\immediate\write18{tex acmart.ins}

\documentclass[acmsmall]{acmart}
\AtBeginDocument{%
  }

\setcopyright{acmlicensed}
\copyrightyear{2018}
\acmYear{2018}
\acmDOI{XXXXXXX.XXXXXXX}
\acmConference[Conference acronym 'XX]{Make sure to enter the correct
  conference title from your rights confirmation email}{June 03--05,
  2018}{Woodstock, NY}
\acmISBN{978-1-4503-XXXX-X/2018/06}

\usepackage{xcolor}
\usepackage{listings}

\lstdefinestyle{pythonstyle}{
    language=Python,
    basicstyle=\ttfamily\small,
    keywordstyle=\color{blue}\bfseries,
    stringstyle=\color{brown},
    commentstyle=\color{gray}\itshape,
    numbers=left,
    numberstyle=\tiny\color{gray},
    stepnumber=1,
    numbersep=8pt,
    backgroundcolor=\color{gray!5},
    frame=single,
    rulecolor=\color{black!30},
    frameround=tttt,
    breaklines=true,
    showstringspaces=false,
    tabsize=4,
    captionpos=b
}

\lstdefinestyle{cobolstyle}{
    language=COBOL,
    basicstyle=\ttfamily\small,
    keywordstyle=\color{blue}\bfseries,
    stringstyle=\color{brown},
    commentstyle=\color{gray}\itshape,
    numbers=left,
    numberstyle=\tiny\color{gray},
    stepnumber=1,
    numbersep=8pt,
    backgroundcolor=\color{gray!5},
    frame=single,
    rulecolor=\color{black!30},
    frameround=tttt,
    breaklines=true,
    showstringspaces=false,
    tabsize=4,
    captionpos=b
}

\lstdefinestyle{cstyle}{
    language=C,
    basicstyle=\ttfamily\small,
    keywordstyle=\color{blue}\bfseries,
    stringstyle=\color{brown},
    commentstyle=\color{gray}\itshape,
    numbers=left,
    numberstyle=\tiny\color{gray},
    stepnumber=1,
    numbersep=8pt,
    backgroundcolor=\color{gray!5},
    frame=single,
    rulecolor=\color{black!30},
    frameround=tttt,
    breaklines=true,
    showstringspaces=false,
    tabsize=4,
    captionpos=b
}

\usepackage{xspace}
\usepackage{siunitx}
\usepackage{multirow}
\usepackage{subcaption}  
\usepackage{tcolorbox}

\begin{document}

\title{SEDCoT: Enhancing LLM-Based COBOL Code Translation via Symbolic Execution and Delta Debugging}

\author{Phillip Entin}
\authornote{Both authors contributed equally to this work.}
\email{phillip.entin@usi.ch}
\orcid{0009-0009-3591-5589}
\affiliation{%
  \institution{University of Lugano}
  \country{Switzerland}
}

\author{Wenchao Gu}
\authornotemark[1]
\authornote{ Corresponding author.}
\email{wenchao.gu@tum.de}
\orcid{0000-0003-3503-8845}
\affiliation{%
  \institution{Technical University of Munich}
  \country{Germany}}

\author{Alexander Knapp}
\email{alexander.knapp@uni-a.de}
\orcid{0000-0002-4050-3249}
\affiliation{%
  \institution{University of Augsburg}
  \country{Germany}
}

\author{Chunyang Chen}
\email{chun-yang.chen@tum.de}
\orcid{0000-0003-2011-9618}
\affiliation{%
 \institution{Technical University of Munich}
 \country{Germany}}

\renewcommand{\shortauthors}{Phillip Entin et al.}

\newcommand{\wcgu}[1]{\textcolor{red}{[#1]}}
\newcommand{\revise}[1]{\textcolor{black}{#1}}
\newcommand{\phen}[1]{\textcolor{orange}{[#1]}}
\newcommand{\tool}{SEDCoT\xspace} 
\newcommand{\code}[1]{\colorbox{gray!10}{\lstinline[basicstyle=\ttfamily]|#1|}}

\newcommand{\summary}[1]{%
\begin{tcolorbox}[
    colback=gray!20,  
    colframe=gray!20,  
    rounded corners, 
    arc=0.8mm,
    boxrule=0.2mm,  
    boxsep=0mm,    
    before skip=5pt,
    after skip=5pt,
]
#1
\end{tcolorbox}
}

\begin{abstract}
COBOL remains critical across banking, insurance, and government infrastructure. However, maintenance is increasingly challenging due to outdated technologies, sparse documentation, and developer retirement, necessitating code translation into modern languages like C. Traditional rule-based transcompilers yield outputs that are difficult to read and maintain, while general-purpose large language models (LLMs) achieve suboptimal correctness because COBOL is a low-resource language with distinct logic patterns. To bridge this gap, we propose \tool, a novel COBOL-to-C translation framework. \tool first leverages LLMs for initial translation, then combines symbolic execution with LLM guidance to generate test suites and iteratively repair semantic discrepancies. Finally, it integrates delta debugging to minimize failing tests into succinct counterexamples, accelerating automated code repair. Evaluating \tool on a public COBOL-to-C dataset demonstrates that it outperforms state-of-the-art baselines by at least 12\% while producing translations with substantially higher readability than rule-based alternatives.

\end{abstract}



\received{20 February 2007}
\received[revised]{12 March 2009}
\received[accepted]{5 June 2009}

\maketitle



\section{Introduction}
Legacy systems remain critical across many sectors, including banking, insurance, and government infrastructure~\cite{weltcobolartikel}. Despite their age, these systems continue to support mission-critical operations — processing \$3 trillion in commerce transactions every day~\cite{Cassel01}. \revise{However, maintaining them is increasingly difficult due to outdated technologies, sparse digital documentation, and the retirement of original developers~\cite{weltcobolartikel, Cassel01}.} As technical debt accumulates, organizations face growing pressure to modernize their infrastructure to improve maintainability, scalability, and compliance~\cite{weltcobolartikel, Cassel01}.

Among legacy technologies, COBOL stands out as particularly entrenched and challenging. Originally developed in the 1960s for business applications~\cite{weltcobolartikel, Cassel01}, COBOL remains widely used today; estimates suggest that more than 220 billion lines of COBOL code are still operational~\cite{reuterscobol}. These programs are deeply embedded in financial systems, mainframes, and government platforms~\cite{weltcobolartikel}.

COBOL’s verbose, data-centric syntax and reliance on legacy constructs make it increasingly incompatible with contemporary development practices. \revise{Translating COBOL into mainstream programming languages therefore requires more than a purely syntactic mapping. COBOL programs commonly employ hierarchical data definitions, atypical control-flow constructs, and tight coupling between data layout and program logic~\cite{iso_cobol_standard};} these characteristics create a substantial semantic gap between COBOL and mainstream programming languages.

\revise{Rule-based tools such as the GnuCOBOL compiler can emit code in mainstream programming languages from COBOL source~\cite{GnuCOBOLHP}. However, because of the significant grammatical and idiomatic differences between COBOL and contemporary languages, the generated code is often difficult for human developers to read and maintain, and typically requires extensive manual refactoring before it is suitable for long-term maintenance. Correctness would indeed be paramount if the translation were like a compilation—"compile, run, and forget." But readability will outweigh "simple" correctness if the translated code has to be worked with in any respect, in particular, if it will have to be maintained as is the usual case in replacing legacy software.}

Driven by this critical need for readability, recent advances in large language models (LLMs) have produced impressive results in program translation\cite{unitrans,Pan_2024,gandhi2024translation,ibrahimzada2025alphatrans,recentllmadvancement}. Unlike rule-based methods, LLM-generated translation code exhibits readability comparable to human-written programs with more natural coding styles and idiomatic expressions. Nevertheless, while LLMs excel in readability, they suffer from a sharp accuracy drop compared with rule-based approaches, and this limitation mainly stems from two key factors. First, COBOL is a low-resource language: far less open-source COBOL code is available for training compared with mainstream programming languages, which limits model exposure to COBOL-specific idioms. Second, COBOL’s grammar and design contain features that do not map cleanly to mainstream programming languages — for example, COBOL’s support for decimal/fixed-point arithmetic and implicit variable initialization at program start~\cite{weltcobolartikel, GnuCOBOLPM} — making it hard for models to learn correct cross-language behaviors from data in other languages. These factors help explain why prior LLM-based COBOL translation efforts~\cite{gandhi2024translation} lag behind LLM performance on translations between mainstream programming languages.

\revise{To address the severe accuracy degradation of LLM-based COBOL translation while preserving readability}, we propose \tool, a novel framework operating in three phases: initial translation, test-case generation, and code repair. First, we prompt LLMs to produce candidate C translations from COBOL. Second, we combine symbolic execution with LLM guidance for test generation: symbolic execution component systematically explores branches and synthesizes inputs to maximize coverage and expose semantic mismatches~\cite{cadar2008klee}, while LLMs increase input diversity. Third, we execute the generated test suite against candidates and supply failing tests to the LLM for automated repair. For complex failures, we apply delta debugging to minimize failing inputs into succinct counterexamples, effectively reducing the search space and helping the LLM localize bugs~\cite{zeller1999yesterday}.

We evaluate \tool on 319 COBOL programs from IBM’s CodeNet dataset, translating them into C due to its prevalence in legacy modernization pipelines. For each program, we synthesize approximately 500 diverse test cases by varying input values, mixing data types, altering whitespace, and injecting non-printable characters. Experimental results show that \tool substantially improves translation accuracy by at least 12\% over state-of-the-art baselines, demonstrating more robust and semantically faithful correctness.

This work makes the following key contributions:
\begin{itemize}
    \item We propose a novel LLM-based code translation framework that integrates initial code translation with translated code repair, thereby improving overall translation accuracy.
    \item We propose a symbolic-execution-based approach to automatically generate test cases with higher line and branch coverage, which are more effective in triggering potential bugs in the initially translated code and guiding the LLM during the repair process.
    \item We propose a delta-debugging-based method to simplify complex test cases which are hard for LLMs to repair, enabling LLMs to more effectively localize buggy code segments.
    \item We conduct a comprehensive evaluation on a public dataset against two state-of-the-art (SOTA) baselines. Experimental results demonstrate that our method significantly outperforms existing SOTA approaches.
\end{itemize}
\section{Background}
\label{sec:background}

\subsection{Symbolic Execution}
\revise{Symbolic execution utilizes constraint solvers to derive inputs that explore distinct program paths~\cite{cadar2008klee}. UTBot industrializes this, employing a hybrid concolic engine combining an enhanced KLEE Symbolic Virtual Machine with smart fuzzing to maximize code coverage while mitigating path explosion~\cite{utbot_symbolic_execution}. Notably, UTBot synthesizes raw KLEE-derived inputs into readable, structured regression test suites based on the Google Test framework, featuring automated method variable generation, assertions, and mock stubs.}



\subsection{Delta Debugging}
When a program exhibits erroneous behavior on a particular input, the input often contains superfluous elements that are not necessary to trigger the fault. Delta debugging, first introduced by Zeller \cite{zeller1999yesterday}, is a widely used technique to automatically minimize such failure-inducing inputs. The core idea is to iteratively partition the input and test whether individual subsets are still capable of reproducing the failure. By systematically discarding irrelevant components, delta debugging converges toward a minimal configuration that preserves the fault-inducing property.

\begin{figure}[ht!]
\centering
\includegraphics[width=9cm]{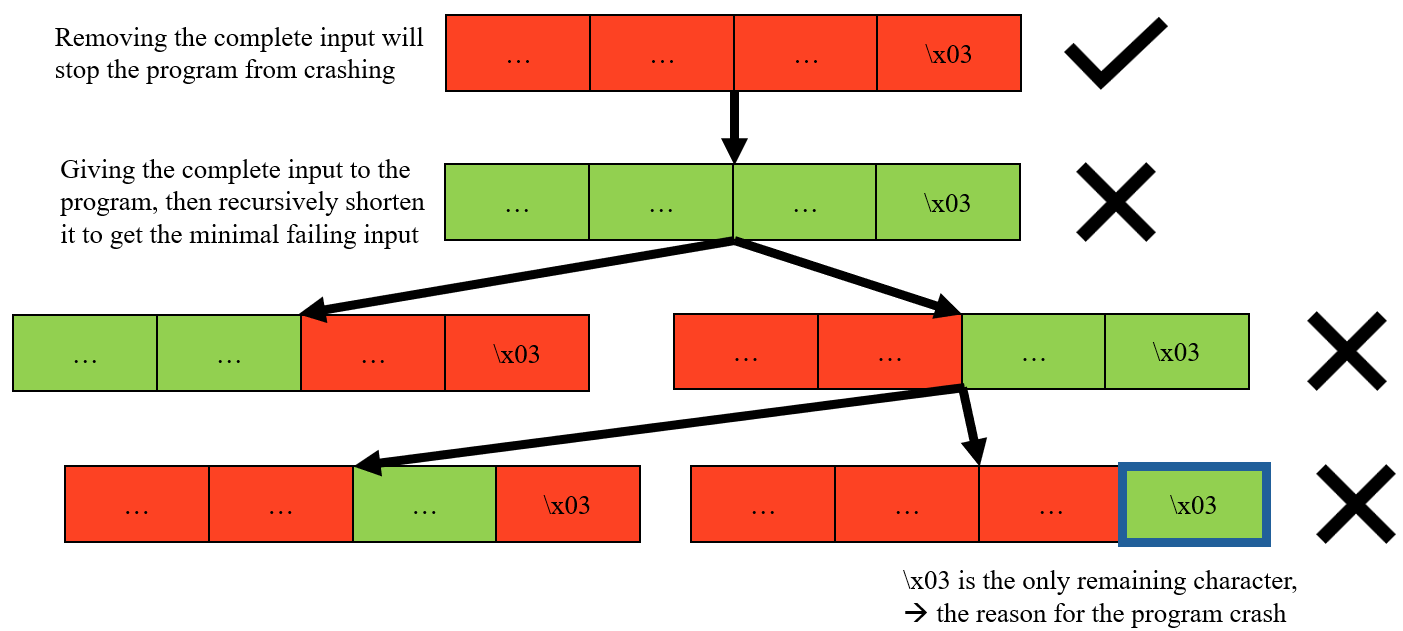}
\caption{Example of delta debugging: a string containing \texttt{\textbackslash x03} is reduced to the minimal failing input \texttt{\textbackslash x03}.}
\label{fig:delta-debugging}
\end{figure}

For example, suppose a program fails when given a string containing ``\texttt{\textbackslash x03}''. Delta debugging may reduce this input to ``\texttt{\textbackslash x03}'' if the shorter prefix alone suffices to reproduce the bug, as illustrated in Figure~\ref{fig:delta-debugging}. In this case, the non-printable character ``\texttt{\textbackslash x03}'' is likely the fault trigger, potentially due to improper handling of non-printable characters in the input parser. Such test case reduction is not only valuable for human developers—by simplifying the debugging process—but also plays a key role in our work: it enables large language models (LLMs) to focus on the essential features of failing cases rather than being distracted by irrelevant context.

\subsection{COBOL Translation}
Open-source solutions like GnuCOBOL enable the compilation and execution of COBOL today~\cite{GnuCOBOLHP} by translating COBOL code into an intermediate C representation~\cite{GnuCOBOLSF}. GnuCOBOL successfully passes 9700 out of 9748 tests in the NIST COBOL 85 test suite~\cite{GnuCOBOLSF}, a benchmark verifying compliance with the COBOL 85 standard~\cite{nistcobol}. While this high coverage demonstrates that GnuCOBOL preserves the semantic correctness of the original programs, its rule-based translation mechanism yields C code that is often difficult for humans to read or maintain. 

For example, a simple COBOL program checking for consecutive doublets is translated into 283 lines of C code, supplemented by two headers of 26 and 59 lines, whereas a human-written C solution requires only about 20 lines. Consequently, we adopt GnuCOBOL as our primary rule-based baseline to compare its readability against the code produced by \tool.

\section{Methodology}
\subsection{Overview}
\begin{figure}[ht]
\centering
\includegraphics[width=0.8\linewidth]{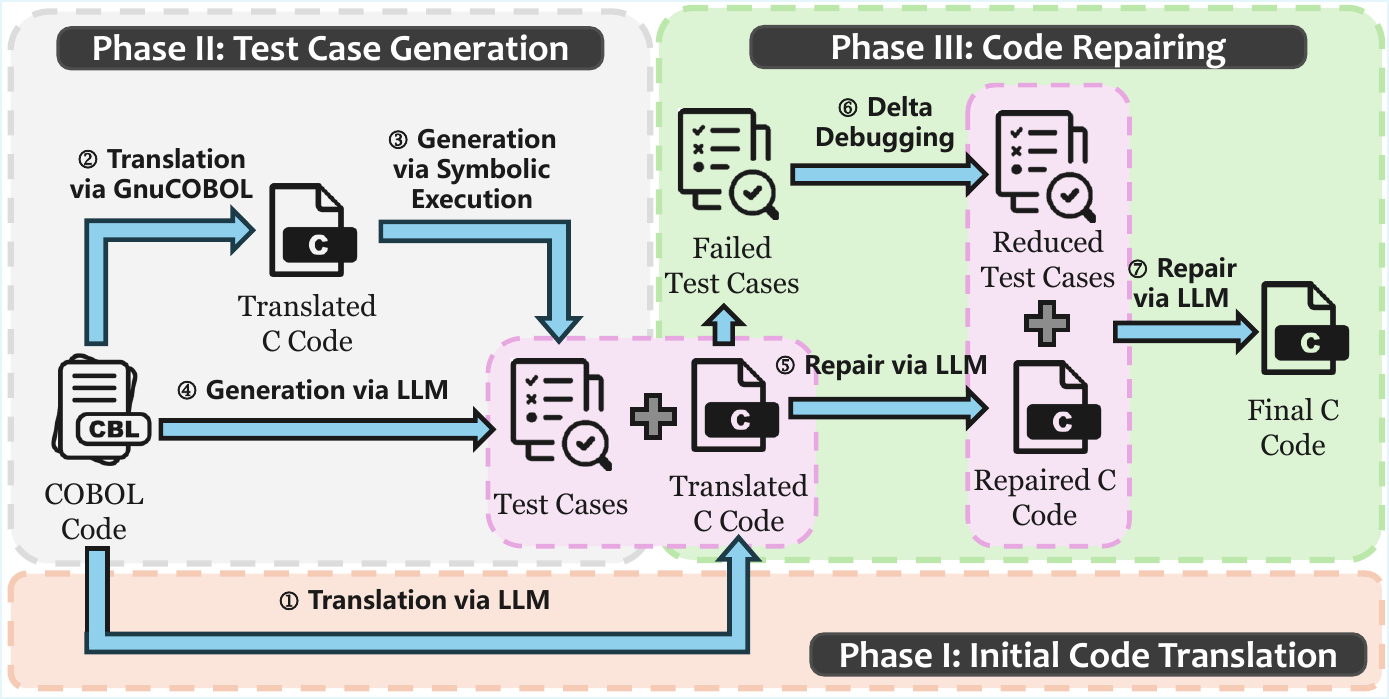}
\caption{Overview of our test-driven COBOL-to-C translation refinement pipeline.}
\label{fig:framework}
\end{figure}

To improve COBOL translation accuracy, we propose a novel approach named \tool, which integrates symbolic execution and delta debugging into test case generation and refinement for code repair during the translation. As illustrated in Figure~\ref{fig:framework}, \tool is organized into three sequential phases: initial code translation, test case generation, and code repair. Notably, symbolic execution is exclusively adopted in the second phase to produce test cases that are subsequently utilized in the third phase, and these two phases run strictly in sequence without forming any iterative loops.

In the initial translation phase, an LLM translates COBOL into the target language to produce candidate code. During test case generation, we construct verification inputs comprising two types: rule-based test cases generated via rule-guided COBOL translation combined with symbolic execution for high code coverage, and LLM-generated test cases. Finally, in the code repair phase, these test cases are executed on the initial translation, and execution feedback is routed to the LLM for iterative error correction. If repeated repair attempts fail, delta debugging simplifies the failing test cases to reduce failure analysis difficulty. This simplified information is then fed back to the LLM to guide subsequent repair and yield the finalized target-language code.

\subsection{Initial Code Translation}
\subsubsection{Translation via LLM}
At the first stage, we leverage LLMs to translate the target COBOL code into the code with target programming language, which serves as the initial translation template for subsequent refinement. The translation process can be formally expressed as:
\begin{equation}
P(C \mid I) = \prod_{t=1}^{T} P(c_t \mid c_{<t}, I),
\end{equation}
where $I$ denotes the input COBOL code, $C$ represents the translated code, $T$ is the sequence length, and $P(\cdot)$ corresponds to the conditional probability distribution over tokens.

Since the initially translated code may contain compilation errors, we attempt to compile the generated code. Let $\mathcal{C}(\cdot)$ denote the compilation function that maps a candidate program $C$ to a diagnostic outcome $E$:
\begin{equation}
E = \mathcal{C}(C),
\end{equation}
where $E = \varnothing$ indicates successful compilation, and $E \neq \varnothing$ corresponds to a set of compilation errors.

If $E \neq \varnothing$, we provide both the erroneous code $C$ and the compiler feedback $E$ to the LLM, enabling iterative refinement. Formally, the refinement step at iteration $k$ can be represented as:
\begin{equation}
C^{(k+1)} = \mathrm{LLM}\big(I, C^{(k)}, E^{(k)}\big),
\end{equation}
where $C^{(k)}$ is the candidate code at iteration $k$, and $E^{(k)}$ is the corresponding compilation feedback. After $K$ iterations, the final code $C^{*}$ is selected as the candidate with the minimal number of compilation errors. Note that the repair process requires executable code; therefore, the subsequent repair procedure will be skipped if $C^{*}$ fails to compile.

\subsection{Test Case Generation}
\subsubsection{Translation via GunCOBOL \& Generation via Symbolic Execution}
To validate the correctness of the translated code and to provide feedback that can assist LLMs in refining the translation, we generate test cases for the original COBOL code. Let $\mathcal{T}(I)$ denote the set of test cases generated for the COBOL program $I$. These test cases are subsequently used for evaluating the translated code $C$.

Most prior approaches rely on LLMs to automatically generate test cases. However, such test cases often suffer from insufficient code coverage, resulting in weak test sets that are easily passed and thus less effective. To address this limitation, we employ symbolic execution techniques, which are capable of systematically exploring program paths to achieve higher code coverage. Formally, let $X = {x_1, x_2, \dots, x_n}$ represent the set of program inputs, which are treated as symbolic variables. The symbolic execution engine explores feasible execution paths $\mathcal{P} = \{p_1, p_2, \dots, p_m\}$, producing a set of test cases $T$ corresponding to each path:
\begin{equation}
T = \{t_i \mid t_i \text{ satisfies path } p_i \in \mathcal{P}\}.
\end{equation}

To the best of our knowledge, no open-source symbolic execution engine currently supports direct test case generation for COBOL. To leverage the proven effectiveness of symbolic execution–based test cases \cite{ibmsymexec}, we adopt the following workaround: the rule-based compiler translates COBOL programs into functionally equivalent intermediate code with target promgramming language. This enables the use of existing symbolic execution engines to generate test cases that capture the behavioral aspects of the original COBOL program.

Formally, let $I_C$ denote the intermediate representation of the COBOL program. The symbolic execution engine $\mathcal{S}$ is then applied to $I_C$ to generate the test cases:
\begin{equation}
T = \mathcal{S}(I_C, X),
\end{equation}
where $X$ denotes the set of symbolic input variables. Each generated test case $t_i \in T$ corresponds to a concrete assignment of $X$ that triggers a unique execution path in $I_C$, thereby providing comprehensive coverage for subsequent evaluation of the translated code.

\subsubsection{Generation via LLM}
Nevertheless, symbolic execution alone may not always generate feasible test cases, as it can struggle with certain program constructs. For instance, loops and nested branches may cause path explosion, while conditions involving non-linear arithmetic can result in constraints that are difficult for SMT solvers to resolve. As a consequence, some execution paths may remain uncovered, leaving no concrete test cases available. To mitigate this limitation, we complement symbolic execution with LLM-generated test cases. \revise{Let $\mathcal{T}_{\mathrm{LLM}}(I)$ denote the set of test cases generated by the LLM from the COBOL source code $I$ (Code translated by GnuCOBOL is lengthy and obscure, making it difficult for LLMs to comprehend and generate test suites), following a procedure similar to the UniTrans approach \cite{unitrans}:}
\begin{equation}
\mathcal{T}_{\mathrm{LLM}}(I) = \mathrm{LLM_{GenerateTests}}(I).
\end{equation}

The final set of test cases for evaluating the translated C code is then obtained by combining the symbolic execution and LLM-generated test cases:
\begin{equation}
T_{\mathrm{final}} = T \cup \mathcal{T}_{\mathrm{LLM}}(I),
\end{equation}
where $T$ is the set of symbolic execution–based test cases defined previously.

Finally, the COBOL program is executed with the inputs in $T_{\mathrm{final}}$, and the corresponding outputs are recorded as the ground truth $O_{\mathrm{COBOL}}$:
\begin{equation}
O_{\mathrm{COBOL}} = \{ o_i \mid o_i = \mathrm{Execute}(I, t_i),  t_i \in T_{\mathrm{final}} \}.
\end{equation}
These ground-truth outputs are used to validate the correctness of the translated code and to provide feedback for subsequent refinement.

\subsection{Code Repairing}
\subsubsection{Repair via LLM}
Once the test cases $T_{\mathrm{final}}$ and a compilable code $C$ are obtained, we first evaluate the generated code against the provided test cases. Let the execution of candidate code $C^{(k)}$ on test case $t_i$ produce output $o_i^{(k)}$, and define the set of failing test cases as
\begin{equation}
\mathcal{E}^{(k)} = \{ t_i \in T_{\mathrm{final}} \mid o_i^{(k)} \neq o_i \},
\end{equation}
where $o_i$ is the corresponding ground-truth output from the COBOL program.

The repair pipeline is initiated to iteratively re-prompt the LLM to correct failing test cases and produce updated translations. At iteration $k$, the LLM receives as input the original COBOL code $I$, the current candidate $C^{(k)}$, and the execution results for only the failing test cases ${ (t_i, o_i^{(k)}, o_i) }_{t_i \in \mathcal{E}^{(k)}}$, following the approach of \cite{Pan_2024}:
\begin{equation}
C^{(k+1)} = \mathrm{LLM_{Repair}}\Big(I, C^{(k)}, { (t_i, o_i^{(k)}, o_i) }_{t_i \in \mathcal{E}^{(k)}} \Big).
\end{equation}

During the repair phase, we additionally verify the output format to ensure that superficially correct values with extraneous spaces, leading zeros, or numeric formatting differences are treated as failures.

A maximum number of repair iterations $K_{\mathrm{max}}$ is enforced. After $K_{\mathrm{max}}$ attempts, the final candidate $C^*$ is selected as the version that passes the largest number of test cases. Let
\begin{equation}
F^{(k)} = \sum_{t_i \in T_{\mathrm{final}}} \mathbf{1}[o_i^{(k)} = o_i]
\end{equation}
denote the total number of passing test cases, then
\begin{equation}
C^* = \arg\max_{0 \le k \le K_{\mathrm{max}}} F^{(k)}.
\end{equation}

If a candidate fails to compile during the repair process, the most recent successfully compiled version is restored and used in subsequent iterations. This procedure ensures that the selected final version is both compilable and achieves the highest overall test-case success.

\subsubsection{Delta Debugging}
When a candidate translation continues to fail certain test cases after the compilation and repair stages, naively re-prompting the LLM often yields diminishing returns, as the remaining failures involve errors that are difficult for the model to identify or reason about. Building on the observation of Yang et~al.~\cite{unitrans} that iterative repairs tend to saturate quickly, we adopt delta debugging to isolate the root causes of these residual errors and provide them to the LLM as explicit repair guidelines.

Let $\mathcal{E}^*$ denote the set of failing test cases remaining after standard repair iterations. For each failing input $t_i \in \mathcal{E}^*$, delta debugging produces a minimal counterexample $t_i^{\min}$ that still reproduces the error:
\begin{equation}
t_i^{\min} = \mathrm{DeltaDebug}(t_i), t_i \in \mathcal{E}^*,
\end{equation}
where $\mathrm{DeltaDebug}(\cdot)$ iteratively shortens and simplifies the input until a minimal reproducible input is obtained \cite{zeller1999yesterday}. The resulting set of minimal counterexamples is
\begin{equation}
\mathcal{E}^{\min} = \{ t_i^{\min} \mid t_i \in \mathcal{E}^* \}.
\end{equation}

\begin{lstlisting}[style=cobolstyle, caption={Minimal COBOL example}, label=lst:foocob]
DATA DIVISION.
...
01  S PIC X(01). 
PROCEDURE DIVISION.
    ACCEPT S
    ACCEPT T
    IF S = T THEN DISPLAY "foo" END-IF
    STOP RUN.
\end{lstlisting}

These minimal inputs often reveal the underlying fault. When $t_i^{\min}$ reduces to an empty string or a single character, the failure is typically caused by COBOL’s implicit initialization semantics (e.g., variables in working storage are automatically initialized in GnuCOBOL \cite{GnuCOBOLPM}). Listing \ref{lst:foocob} illustrates a minimal example of this issue: the program requests two string inputs from the user and prints \texttt{foo} if they are equal. However, if the user provides no input, the program still prints \texttt{foo} due to GnuCOBOL’s automatic variable initialization. Longer minimal inputs containing unusual whitespace or control characters typically indicate unsafe input handling or incorrect assumptions regarding input length.

To avoid overwhelming the LLM with redundant examples, we normalize and deduplicate failing test cases. Let $\mathrm{Canonical}(t_i^{\min})$ denote the normalized form of $t_i^{\min}$ (e.g., replacing ASCII codes below 32 with a dot). The final set of unique minimal counterexamples is
\begin{equation}
\mathcal{E}^{\mathrm{canon}} = \{ \mathrm{Canonical}(t_i^{\min}) \mid t_i^{\min} \in \mathcal{E}^{\min} \text{ and duplicates removed} \}.
\end{equation}

By providing only the unique, minimal failing inputs, this delta-debugging reduction step supplies the LLM with precise, high-level guidance, rather than a large number of opaque examples. \revise{Consequently, we achieve better repair outcomes in the final stage without extra LLM calls. Delta debugging helps LLMs precisely identify errors and finish fixes efficiently, avoiding repeated model invocations for iterative reasoning and trial repairs.}

However, an early application of delta debugging may prematurely constrain the search space of LLMs and bias the subsequent repair steps toward overly specific fixes. Suppose that delta debugging adopts a test oracle that only discriminates between \emph{fail} and \emph{pass} outcomes. In this scenario, delta debugging performs input minimization merely based on the preservation of failure behavior, with no consideration for the distinct underlying causes of failures. A representative example is given by the pseudocode in Listing \ref{lst:dd-ex}:

\begin{lstlisting}[style=pythonstyle, caption={Minimal DD Example}, label=lst:dd-ex]
if input == "abc": crash
elif input == "ab": crash
elif input == "a": return expected
\end{lstlisting}

When the input \texttt{"abc"} is given, the program crashes. Since the input \texttt{"ab"} also triggers a crash, delta debugging reduces the original input to \texttt{"ab"} as the minimal failure-inducing input. As a result, the bug in the branch corresponding to \texttt{"abc"} becomes unobservable.

To address this issue, \tool first attempts to fix the bug using the full input in the initial several rounds, and only employs delta debugging in the final round to perform input minimization.

\subsubsection{Repair via LLM} Once the reduced test cases $\mathcal{E}^{\mathrm{canon}}$ are generated, their corresponding execution results ${o_i^{\min}}$, together with the current translated code $C$, are used to guide the final repair step:
\begin{equation}
C^{*} = \mathrm{LLM_{FinalRepair}}\Big(C, { (t_i^{\min}, o_i^{\min}, o_i) \mid t_i^{\min} \in \mathcal{E}^{\mathrm{canon}}} \Big),
\end{equation}

\noindent where $o_i$ denotes the ground-truth output produced by the original COBOL program. This formulation ensures that the LLM focuses specifically on the remaining minimal failing inputs, leveraging precise execution feedback to generate the final repaired code $C^*$. In addition, we design instructions that explicitly encourage the LLM to improve robustness in input handling and variable initialization—issues that frequently arise in COBOL translation.
\section{Experimental Settings}
In this section, we introduce the datasets, baselines and LLMs being evaluated in the experiments.

\subsection{Dataset}

\begin{table}[htbp]
  \centering
  \begin{minipage}[t]{0.35\textwidth}
    \centering
    \small
    \caption{Statistics for Lines of Code}
    \begin{tabular}{lllll}
      \toprule
      & Min & Median & Mean & Max \\
      \midrule
      LOC & 9.0 & 32.0 & 37.2 & 210.0 \\
      \bottomrule
    \end{tabular}
    \label{tab:loc}
  \end{minipage}
  \hfill 
  \begin{minipage}[t]{0.6\textwidth}
    \centering
    \small
    \caption{Percentage Distribution of different Lines}
    \begin{tabular}{lllll}
      \toprule
      & \textless 20 & 20 $\sim$ 50 & 51 $\sim$ 100 & \textgreater 100 \\
      \midrule
      Num & 58 (18.2\%) & 198(62.1\%) & 54(16.9\%) & 9 (2.8\%) \\
      \bottomrule
    \end{tabular}
    \label{tab:distribution}
  \end{minipage}

\end{table}

Building upon the work of Gandhi et al.~\cite{gandhi2024translation}, we adopt the IBM CodeNet dataset and extract 322 COBOL programs that not only compile successfully but also pass their associated sample tests (marked as "accepted") for our evaluation. \revise{However, IBM dataset does not provide dialect version information, which leads to compilation and reproducibility issues. We therefore require all programs to be reproducibly compilable using a controlled, uniform toolchain (GnuCOBOL); under this stricter and experimentally necessary criterion, 319 programs are retained. GnuCOBOL is a reliable open-source COBOL-to-C translation tool to convert COBOL programs into C code. We selected GnuCOBOL over alternative dialects for two primary reasons. First, Project CodeNet metadata explicitly designates "OpenCOBOL 1.1.0"—the direct predecessor of GnuCOBOL—as the source environment for all submissions. Second, migrating this legacy source code to alternative dialects would invariably introduce severe compilation discrepancies.} As documented in its official manual~\cite{GnuCOBOLSF}, GnuCOBOL achieves a 98.3\% pass rate on the NIST COBOL85 test suite, with only the "Advanced facility" subsets of the "CM - COMMUNICATION SECTION tests", "DB - Debugging facilities tests", and "OB - Obsolete facilities tests" remaining untested. Despite not achieving 100\% accuracy, its output is still regarded as the ground-truth translated code for this study. \revise{Table~\ref{tab:loc} and Table~\ref{tab:distribution} shows the statistical results about the dataset.} For system execution tools, KLEE~\cite{cadar2008klee} is adopted for rule-based test case generation.

We attempted to validate our method against other mainstream languages. However, our framework requires reliable baseline transcompilers to support subsequent test generation and behavior comparison. Currently, robust open-source translation tools for other target languages remain unavailable: for instance, translating COBOL to Java via an open-source tool~\cite{cob2j} yielded only 11 successful translations, an insufficient sample size for meaningful statistical evaluation. Consequently, we limit our target language to C for all experiments in this study.

\subsection{Evaluation}

\revise{Because the dataset's original test cases are overly simplistic, we adopt a differential testing paradigm using a held-out golden test suite. We set the generated input volume to 500 per program to balance computational overhead and anomaly detection. Mutating from each benchmark's original input, we generate these 500 test cases evenly across five strategies (100 samples each): type-preserving shuffling, mixed-type shuffling, whitespace perturbation, variable-length truncation with non-printable characters, and variable-length truncation with mixed character replacement. This test suite achieves 88.42\% statement coverage and 52.97\% branch coverage. Crucially, no translated program passes the entire suite, demonstrating its comprehensiveness in uncovering behavioral discrepancies.}

\revise{Since native COBOL code cannot be directly executed on modern hardware, we leverage GnuCOBOL to compile and run the original COBOL programs, and take its execution behavior as the ground-truth oracle. We then compare the runtime outputs of LLM-translated programs against this GnuCOBOL-based reference to rigorously examine the behavioral equivalence and robustness of translation results. In our differential testing pipeline, GnuCOBOL therefore acts as the authoritative execution standard. Finally, we quantify the performance of all methods by computing the average test case pass ratio over the 500 perturbed inputs per program.}

\subsection{Baselines and LLMs evaluated}
We evaluate \tool on four representative large language models that vary in architecture, parameter scale, and context-window size. This selection enables a systematic assessment of \tool's performance across different generative systems. \revise{All LLMs were deployed and executed on cloud platforms. Specifically, GPT-4o-mini ran on OpenAI's service infrastructure, and other models were accessed via diverse providers supported by OpenRouter~\cite{openrouter}.}

\begin{itemize}
    \item \emph{Gemma-3-27b-it} (Google)~\cite{OpenRouterGemma}: an instruction-tuned model with 27 billion parameters and a context window of 128,000 tokens.
    \item \emph{Llama-3.3-70b-instruct} (Meta)~\cite{OpenRouterLlama}: an instruction-tuned model with 70 billion parameters and a context window of 131,000 tokens.
    \item \emph{Qwen-2.5-coder-32b-instruct} (Qwen)~\cite{OpenRouterQwen}: a coder-oriented model with 32 billion parameters and a context window of 32,000 tokens.
    \item \emph{GPT-4o-mini-2024-07-18} (OpenAI)~\cite{4omini}: a general-purpose model with undisclosed parameter count; its context window is accessed via the official OpenAI API.
\end{itemize}

\revise{In addition, we evaluate the efficacy of \tool by comparing it against one rule-based approach and two state-of-the-art LLM-based baseline methods:}

\begin{itemize}
    \item \revise{\textbf{TinyCOBOL}: an early open-source COBOL 85 compiler that directly translates COBOL source code into x86 assembly language.}
    \item \textbf{UniTrans}~\cite{unitrans}: an iterative repair procedure driven solely by LLM-generated tests, originally designed for program translation tasks in modern programming languages such as Java, Python and C++. \revise{As a representative general LLM-based code translation framework, it is adopted to verify the applicability of such mainstream solutions to legacy code translation scenarios. We retain its full pipeline and only adjust prompts and compiler settings to support COBOL-to-C translation. For fair comparison, we limit its iterative repair to three rounds, consistent with \tool.}
    \item \textbf{High-Resource Java Refinement (HRJR)}~\cite{gandhi2024translation}: a three-step logic–readability–logic refinement sequence proposed for COBOL-to-Java translation. \revise{We retained the complete workflow of HRJR, and only adjusted the LLM prompt to generate C code instead of Java, while replacing the compiler accordingly.}
\end{itemize}


\subsection{Implementation Details}
In our experiments, the temperature of all LLMs was set to zero to eliminate sampling variance~\cite{renze2024effect}. To mitigate the high time complexity of symbolic execution, we imposed maximum allowances for each COBOL program. Specifically, KLEE was configured to explore up to 80 million expressions, restricted to an overall time limit of 20 minutes, 2 minutes per state, and 5 minutes for coverage measurement. Test generation via KLEE was conducted on an Intel i5-11300H laptop (16GB RAM) over 12 hours, successfully producing test suites for 292 programs with an average runtime of 2.5 minutes. All prompts used in our method are provided in the replication package's Appendix.

\begin{table}[ht]
    \centering
    \small
    \caption{Statistics of Test Cases Generated by Symbolic Execution and LLMs.}
    \begin{tabular}{llllll}
    \toprule
    & Symbolic & Gemma-3 & GPT-4o-mini & Llama-3.3 & Qwen-2.5-coder \\
    \midrule
    Mean & 15.25 & 15.82 & 14.06 & 7.22 & 35.32 \\
    Median & 17.00 & 13.00 & 8.00 & 7.00 & 10.00 \\
    Min & 1.00  & 1.00 & 2.00 &  2.00 & 2.00 \\
    Max & 144.00 & 28.00  & 1001.00 & 30.00 & 1001.00 \\
    Succeed Cases & 292 & 293 & 308 & 303 & 303 \\
    \bottomrule
    \end{tabular}
    \label{tab:test_stat}
\end{table}

\begin{table}[ht]
    \centering
    \scriptsize
    \setlength{\tabcolsep}{3pt} 
    \caption{Overlap of Test Cases Generated by Different Approaches.}
    \begin{tabular}{lllllllllllll}
    \toprule
    & Sym\&Gem & Sym\&GPT & Sym\&Lla & Sym\&Qwe & Gem\&GPT & Gem\&Lla & Gem\&Qwe & GPT\&Lla & GPT\&Qwe &  Lla\&Qwe \\
    \midrule
    Overlap & 270 (84.6\%) & 282 (88.4\%) & 277 (86.8\%) & 276 (86.5\%) & 284 (89.0\%) & 279 (87.5\%) & 280 (87.8\%) & 294 (92.2\%) & 293 (91.8\%) & 290 (90.9\%) \\
    \bottomrule
    \end{tabular}
    \label{tab:overlap}
\end{table}

Table~\ref{tab:test_stat} summarizes the statistics of the test cases generated by both symbolic execution and LLMs. Here, “Succeed Cases” refers to the number of COBOL programs for which test cases were successfully generated. Since program complexity varies significantly, the number of test cases required to achieve high line and branch coverage also differs widely. Therefore, we did not impose an upper bound on the number of test cases, allowing both symbolic execution and LLM-based methods to generate cases until they naturally terminated. \revise{Table~\ref{tab:overlap} summarizes the number of overlapping test cases generated by different approaches. No test cases are uniquely generated by either one LLM alone or symbolic execution. In addition, all experimental results are reported using the 319 test cases. The repair stage is skipped if no test cases are provided for repair, yet the translated code is still evaluated against the golden test suite.}

\revise{For experimental consistency, we strictly follow the original settings of each baseline: HRJR adopts the native test cases from CodeNet for program repair; UniTrans leverages LLM-generated test cases to conduct repair as originally designed. In terms of our proposed SEDCoT, we employ both LLM-generated and symbolic execution-derived test cases for program repair in RQ1. By contrast, selective combinations of these two types of test cases are adopted in RQ2 and RQ4 to satisfy the requirements of ablation studies.}

In general, a large number of repair retries may lead to diminishing performance~\cite{unitrans}. For example, Gandhi et al.~\cite{gandhi2024translation} limit the maximum number of repair iterations to three in their COBOL translation approach. To ensure consistency with prior work and enable a fair performance comparison, we similarly set the total number of repair attempts per test suite to three. Concretely, in the first code repair stage (Step 5 of Phase 3), LLMs were allowed up to two retries. In the second code repair stage (Step 7), the maximum number of retries was set to one; however, in the event of compilation failures, we allowed up to two additional attempts solely to address compilation errors.

\section{Evaluation}
\label{sec:evaluation}
We aim to answer the following research questions (RQs):
\begin{itemize}
\item \textbf{RQ1:} How effective is \tool in translating COBOL programs compared with SOTA baseline methods?
\item \textbf{RQ2:} To what extent do the individual components of \tool to the overall translation quality?
\item \textbf{RQ3:} Can symbolic execution test cases reveal more latent bugs than LLM test cases?
\item \textbf{RQ4:} How does the number of repair attempts affect the performance of \tool?
\item \textbf{RQ5:} How does the readability of code produced by \tool compare with that of rule-based translation approaches?
\end{itemize}

\subsection{RQ1: How effective is \tool in translating COBOL programs compared with SOTA baseline methods?}



\begin{table}[ht]
\centering
\small
\setlength\tabcolsep{8pt}
\caption{Performance comparison of different approaches with various LLMs (best performance in bold).}
\label{tab:overall}
\begin{tabular}{lllll}
\toprule
& Gemma-3 & GPT-4o-mini & Llama-3.3 & Qwen-2.5-coder \\
 \midrule
Vanilla LLM & 0.287 & 0.393 & 0.338 & 0.345\\
HRJR & 0.455 &  0.380 & 0.502 & 0.504 \\
UniTrans & 0.516 & 0.440 & 0.514 & 0.531 \\
\tool & \textbf{0.579 ($\uparrow$ 12.2\%)} & \textbf{0.718 ($\uparrow$ 63.2\%)} & \textbf{0.607 ($\uparrow$ 18.1\%)} & \textbf{0.668 ($\uparrow$ 25.8\%)}\\
\midrule
TinyCOBOL & 0.207 & 0.207 & 0.207 & 0.207 \\
GnuCOBOL & 1.000 & 1.000 & 1.000 & 1.000 \\
\bottomrule
\end{tabular}
\end{table}

\begin{table}[ht]
\centering
\small
\setlength\tabcolsep{6pt}
\caption{Status of Translated Codes at Different Phases}
\label{tab:detail}
\begin{tabular}{llllll}
\toprule
& & Gemma-3 & GPT-4o-mini & Llama-3.3 & Qwen-2.5-coder \\
 \midrule
\multirow{3}{*}{Compile-error} & Vanilla LLM & 1.3\% (4/319) & 7.2\% (23/319) & 12.2\% (39/319) & 14.4\% (46/319)\\
& HRJR & 4.7\% (15/319) & 15.4\% (49/319) & 16.6\% (53/319) & 17.2\% (55/319)	 \\
& UniTrans & 1.3\% (4/319) & 7.2\% (23/319) & 12.2\% (39/319) & 14.4\% (46/319) \\
& \tool & 2.8\% (9/319) & 0.9\% (3/319) & 12.9\% (41/319) & 6.0\% (19/319)\\
 \midrule
\multirow{3}{*}{Failed} & Vanilla LLM & 91.1\% (275/302) & 87.4\% (270/309) & 88.6\% (242/273) & 85.8\% (254/296)\\
& HRJR & 59.2\% (180/304) & 25.2\% (68/270) & 65.8\% (175/266) & 85.6\% (226/264) \\
& UniTrans & 83.8\% (264/315) & 19.6\% (58/296) & 82.5\% (231/280) & 83.2\% (227/273) \\
& \tool & 81.6\% (253/310) & 77.8\% (246/316) & 73.0\% (203/278) & 74.7\% (224/300)\\
 \midrule
\multirow{3}{*}{Repaired} & HRJR & 2.2\% (4/180) & 0.0\% (0/68) & 1.5\% (4/175) & 1.9\% (5/226) \\
& UniTrans & 1.9\% (5/264) & 1.7\% (1/58) & 2.6\% (6/231) & 2.6\% (6/227) \\
& \tool & 12.6\% (32/253) & 26.4\% (65/246) & 27.6\% (56/203) & 25.9\% (58/224)	\\
\bottomrule
\end{tabular}
\end{table}

Table~\ref{tab:overall} compares the overall performance of \tool with the baselines across different LLMs. \revise{Table~\ref{tab:detail} presents detailed status statistics of translated codes across different stages, including uncompilable programs, compilable codes failing generated test suites, and successfully repaired codes. Notably, no programs can fully pass the golden test suites.}  First, we find that simply adopting LLMs for COBOL code translation yields low accuracy, below 40\%. In contrast, our proposed approach significantly improves accuracy and outperforms state-of-the-art baselines. Specifically, \tool achieves at least a 12.2\% improvement and approximately 30\% average improvement across all LLMs. Furthermore, the relative performance improvement of \tool depends on the base LLM’s translation capability: the stronger the base LLM, the larger the potential improvement provided by \tool. For instance, Gemma-3, which exhibits the lowest baseline performance, gains around 12\% improvement with \tool, whereas GPT-4o-mini, the best-performing LLM, sees a relative improvement exceeding 60\%. \revise{Given that GnuCOBOL serves as the ground-truth oracle in our experimental setup, its performance is defined as the baseline (100\%) for behavioral consistency. Notably, TinyCOBOL exhibits unsatisfactory performance due to severe compatibility limitations. It generates 32-bit assembly code that demands a dedicated 32-bit runtime environment. Although we performed syntactic adaptation to align the assembly output with our GCC compilation configuration and boosted the initial compilation success rate, 206 out of 319 programs still failed to compile owing to syntax discrepancies. As TinyCOBOL stopped receiving updates in January 2011, such compilation failures stem primarily from its inability to support modern COBOL dialects and language variants contained in the CodeNet dataset.}

Interestingly, using LLMs to refactor \texttt{GnuCOBOL}-translated code (denoted as LLM Refactor) yields significantly lower accuracy than direct LLM translation. This occurs because rule-based translated code differs substantially from human-written programs in structure and readability. Lacking conventional coding styles and logical clarity, such translated code is not only opaque to human developers but also prevents LLMs from performing effective refactoring.

\summary{\textbf{Summary 1:} Although LLMs excel at generating syntactically correct, compilation-error-free code, functional correctness is not guaranteed. While \tool significantly boosts LLM translation performance, this improvement is more pronounced when applied to models with stronger baseline translation capabilities.}

\subsection{RQ2: To what extent do the individual components of \tool to the overall translation quality?}

To evaluate the contribution of each component to the overall performance, we conduct a comprehensive ablation study by selectively removing individual components and examining the resulting performance changes. \revise{Specifically, we construct three variants to evaluate individual component contributions: $\rm \tool_{LLMTest}$, $\rm \tool_{SymTest}$, and $\rm \tool_{w/oDelta}$. $\rm \tool_{LLMTest}$ isolates the impact of LLM-generated test cases during the repair stage by removing the symbolic execution component from Phase II. $\rm \tool_{SymTest}$ exclusively utilizes test cases produced by symbolic execution tools, omitting the LLM-based test generation in Phase II. $\rm \tool_{w/oDelta}$ disables the delta debugging process in Phase III, meaning that failed test cases are fed directly into the final repair round without any simplification.Furthermore, since delta debugging inherently requires a foundational test suite to operate, and isolating either pure symbolic-based or pure LLM-based testing has already been extensively explored in literature, we omit further combinations of these two-component ablations.}

\begin{table}[ht]
\centering
\small
\setlength\tabcolsep{8pt}
\caption{Contribution of Different Components in \tool.}
\label{tab:ablation}
\begin{tabular}{lllll}
\toprule
& Gemma-3 & GPT-4o-mini & Llama-3.3 & Qwen-2.5-coder \\
\midrule
$\rm \tool_{LLMTest}$ & 0.488 & 0.547 & 0.513 & 0.503 \\
$\rm \tool_{SymTest}$ & \textbf{0.601 ($\uparrow$ 3.8\%)} & 0.663 & 0.594 & 0.630 \\
$\rm \tool_{w/oDelta}$ & 0.475 & 0.586 & 0.520 & 0.568 \\
\tool & 0.579 & \textbf{0.718 ($\uparrow$ 8.3\%)} & \textbf{0.607 ($\uparrow$ 2.2\%)} & \textbf{0.668 ($\uparrow$ 6.0\%)} \\
\bottomrule
\end{tabular}
\end{table}

Table~\ref{tab:ablation} presents the performance of these variants. First, we observe that \tool consistently achieves the best performance across most LLMs, demonstrating the effectiveness of combining test cases generated by both symbolic execution tools and LLMs, along with delta debugging. Furthermore, by comparing $\rm \tool_{LLMTest}$ and $\rm \tool_{SymTest}$, we find that the variant using symbolic execution–generated test cases significantly outperforms the one relying solely on LLM-generated test cases. This is because symbolic execution can produce test cases covering corner cases, which are more likely to expose bugs and thus provide stronger guidance for LLMs to identify and repair faulty code. \revise{For Gemma-3, symbolic-execution-generated test cases alone yield better performance than combined test sets, further validating this conclusion.} Finally, comparing $\rm \tool_{w/oDelta}$ with \tool highlights the importance of delta debugging: simplifying complex failing test cases that could not be repaired in earlier rounds makes them easier for LLMs to interpret, thereby increasing the likelihood of successful repair in the final round.

\summary{\textbf{Summary 2:} Combining the test cases generated by symbolic execution tools and LLMs, and applying delta debugging to simplify those test cases that are difficult to repair, can significantly improve overall performance. In particular, the use of test cases generated by symbolic execution tools, together with delta debugging, makes a substantial contribution to the effectiveness of our approach.}

\subsection{RQ3: Can symbolic execution test cases reveal more latent bugs than LLM test cases?}

\begin{table}[ht]
    \centering
    \small
    \caption{Statistics of code coverage.}
    \label{tab:coverage}
    \begin{tabular}{llllll}
        \toprule
         Avg. coverage & Symbolic & Gemma-3 & GPT-4o-mini & Llama-3.3 & Qwen-2.5-coder  \\
        \midrule
        Line & 0.852 &0.884 & 0.879 & 0.883 & 0.876 \\
        Branch & 0.469 & 0.524 & 0.515 & 0.525 & 0.522 \\
        \bottomrule
    \end{tabular}
\end{table}

To better understand if test cases generated by symbolic execution tools can more effectively assist LLMs in repairing translated C code, we first compare the code coverage achieved by test cases from both symbolic execution tools and LLMs, as shown in Table~\ref{tab:coverage}. \revise{Coverage was calculated based on the number of programs which can successfully generated test cases for each model. For example, ChatGPT’s average coverage was computed over 308 programs, Gemma’s over 293 programs, and so forth.} Interestingly, the results show no significant difference in coverage between the two approaches. In some cases, LLM-generated test cases even achieve slightly higher line and branch coverage than those produced by symbolic execution.

At first glance, this observation seems contradictory. However, the key lies in the limitations of traditional coverage metrics such as line and branch coverage. These metrics are inherently syntactic: they only measure whether certain statements or branches have been executed, without capturing how test inputs interact with program semantics or whether they are capable of exposing faulty behavior. Symbolic execution, in contrast, tends to generate inputs that exercise edge cases—for example, non-printable characters, malformed data, or extreme boundary values. Such inputs are more likely to trigger faults that remain undetected under typical, syntactically valid but semantically ordinary test cases produced by LLMs.

For instance, consider the classic buffer overflow scenario caused by using 
\texttt{scanf("\%s", ...)} without proper bounds checking, as shown in Listing \ref{lst:example2}:

\begin{lstlisting}[style=cstyle,caption={Unsafe input handling vulnerable to non-printable or long inputs.}, label={lst:example2}]
char buf[8];
scanf("%s", buf);   // unsafe: no length limit
printf("Input: %s\n", buf);
\end{lstlisting}

An LLM may generate a syntactically correct translation of the original COBOL code along with reasonable test inputs (e.g., \texttt{"hello"}) that appear valid. Symbolic execution, however, can systematically produce malformed or non-printable inputs such as \texttt{"\textbackslash x01\textbackslash x02AAAAAAA"}, which can overflow the buffer or disrupt downstream functions. While non-printable characters alone may not immediately crash the program, they can be interpreted as control characters in \texttt{printf("Input: \%s\textbackslash
n", buf);} or, if a null byte (\texttt{\textbackslash x00}) occurs early in the input, prematurely terminate the string, causing subsequent logic to make invalid assumptions.

From a coverage perspective, both the LLM- and symbolic execution-generated test suites may reach the same statements, giving the impression of equivalent coverage. Yet, only the symbolic execution inputs expose the underlying vulnerability, thereby revealing latent bugs that LLM-generated test cases may fail to detect.

\begin{figure}[ht]
\centering
\includegraphics[width=\linewidth]{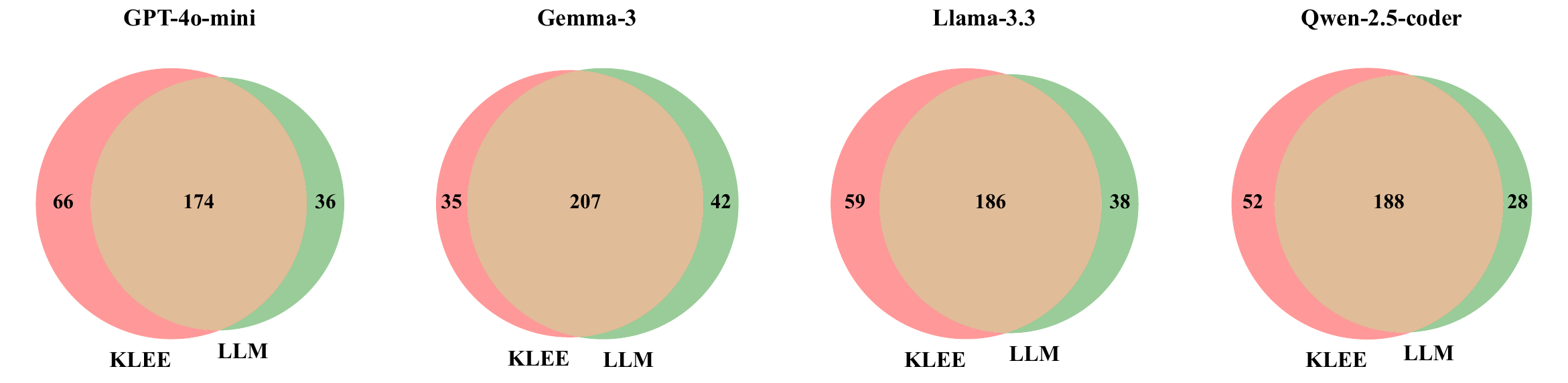}
\caption{Comparison of Bug-Triggering Effectiveness: Symbolic Execution vs. LLM-Generated Test Cases.}
\label{fig:trigger}
\end{figure}

Furthermore, as illustrated in Figure~\ref{fig:trigger}, which shows the number of initially translated programs that failed testing using test cases generated by symbolic execution tools or LLMs, although many bugs can indeed be triggered by test cases from both sources, symbolic execution–based test cases expose a greater number of unique bugs overall. This advantage arises from their ability to systematically generate corner cases that LLM-based test generation often overlooks, particularly for translated programs under minor input variations.

\summary{\textbf{Summary 3:} Although the test cases generated by symbolic execution tools and those produced by LLMs achieve comparable code coverage, the former are able to capture more corner cases. This makes them more effective in guiding LLMs to repair the translated code. }

\subsection{RQ4: How does the number of repair attempts affect the performance of \tool?}

\begin{table}[ht]
\centering
\small
\caption{Impact of repair attempt numbers on the performance of different variants across LLMs (Higher scores denote better performance).}
\label{tab:repair_number}
\begingroup
\setlength{\tabcolsep}{7pt}\small
\begin{tabular}{llllll}
\toprule
& Stage & Gemma-3 & GPT-4o-mini & Llama-3.3 & Qwen-2.5-coder \\
\midrule
\multirow{4}{*}{$\rm \tool_{LLMTest}$}
 & CEF & 0.315 & 0.414 & 0.384 & 0.349 \\
 & Repair-1 & 0.382 ($\uparrow$ 21.3\%) & 0.472 ($\uparrow$ 14.0\%) & 0.415 ($\uparrow$ 8.1\%) & 0.400 ($\uparrow$ 14.6\%) \\
 & Repair-2 & 0.385 ($\uparrow$ 0.8\%) & 0.476 ($\uparrow$ 0.8\%) & 0.436 ($\uparrow$ 5.1\%) & 0.423 ($\uparrow$ 5.8\%) \\
 & Repair-3 & 0.488 ($\uparrow$ 26.8\%) & 0.547 ($\uparrow$ 14.9\%) & 0.513 ($\uparrow$ 17.7\%) & 0.503 ($\uparrow$ 18.9\%) \\
\midrule
\multirow{4}{*}{$\rm \tool_{SymTest}$}
 & CEF & 0.315 & 0.414 & 0.384 & 0.349 \\
 & Repair-1 & 0.483 ($\uparrow$ 53.3\%) & 0.558 ($\uparrow$ 34.8\%) & 0.467 ($\uparrow$ 21.6\%) & 0.478 ($\uparrow$ 37.0\%) \\
 & Repair-2 & 0.486 ($\uparrow$ 0.6\%) & 0.565 ($\uparrow$ 1.3\%) & 0.514 ($\uparrow$ 10.1\%) & 0.499 ($\uparrow$ 4.4\%) \\
 & Repair-3 & 0.601 ($\uparrow$ 23.7\%) & 0.663 ($\uparrow$ 17.3\%) & 0.594 ($\uparrow$ 15.6\%) & 0.630 ($\uparrow$ 26.3\%) \\
\midrule
\multirow{4}{*}{$\rm \tool_{w/oDelta}$}
 & CEF & 0.315 & 0.414 & 0.384 & 0.349 \\
 & Repair-1 & 0.460 ($\uparrow$ 46.0\%) & 0.590 ($\uparrow$ 42.5\%) & 0.472 ($\uparrow$ 22.9\%) & 0.479 ($\uparrow$ 37.2\%) \\
 & Repair-2 & 0.471 ($\uparrow$ 2.4\%) & 0.603 ($\uparrow$ 2.2\%) & 0.510 ($\uparrow$ 8.1\%) & 0.525 ($\uparrow$ 9.6\%) \\
 & Repair-3 & 0.475 ($\uparrow$ 0.8\%) & 0.586 ($\downarrow$ 2.8\%) & 0.520 ($\uparrow$ 2.0\%) & 0.568 ($\uparrow$ 8.2\%) \\
\midrule
\multirow{4}{*}{\tool}
 & CEF & 0.315 & 0.414 & 0.384 & 0.349 \\
 & Repair-1 & 0.460 ($\uparrow$ 46.0\%) & 0.590 ($\uparrow$ 42.5\%) & 0.472 ($\uparrow$ 22.9\%) & 0.479 ($\uparrow$ 37.2\%) \\
 & Repair-2 & 0.471 ($\uparrow$ 2.4\%) & 0.603 ($\uparrow$ 2.2\%) & 0.510 ($\uparrow$ 8.1\%) & 0.525 ($\uparrow$ 9.6\%) \\
 & Repair-3 & 0.579 ($\uparrow$ 22.9\%) & 0.718 ($\uparrow$ 19.1\%) & 0.607 ($\uparrow$ 19.0\%) & 0.668 ($\uparrow$ 27.2\%) \\
\bottomrule
\end{tabular}
\endgroup
\end{table}

To investigate repair iterations, we evaluate translation accuracy across successive attempts, denoted in Table~\ref{tab:repair_number} as CEF (compilation error fixing via LLMs with up to two attempts) and Repair-N (the $N$-th repair round with test cases). Bracketed percentages indicate relative improvements over the preceding step. Comparing CEF and Repair-1 reveals that initial accuracy remains low after compilation fixes, whereas test-case feedback in the first attempt substantially boosts performance. However, this gain diminishes in Repair-2. By Repair-3, improvement for $\rm \tool_{w/oDelta}$ stagnates, and for certain models like GPT-4o-mini, accuracy slightly degrades.

Conversely, delta debugging successfully sustains the LLM's repair capability for complex cases, as evidenced by comparing Repair-2 and Repair-3 across $\rm \tool_{LLMTest}$, $\rm \tool_{SymTest}$, and \tool. Regardless of whether test suites are synthesized via symbolic execution, LLMs, or their combination, delta debugging consistently enhances repair effectiveness. 

Listing~\ref{lst:example1} illustrates this by isolating a subtle input-related bug. The faulty translation uses \texttt{scanf("\%s", ...)} to buffer input, followed by \texttt{sscanf(...)} to parse two integers. Because \texttt{scanf("\%s")} terminates at the first whitespace, the buffer captures only a single token, causing \texttt{sscanf} to leave the second variable (\texttt{y}) uninitialized. While this defect remains masked under seemingly valid inputs like \texttt{"10 2"} due to residual memory contents, delta debugging systematically minimizes the failure-inducing input to \texttt{"1"}. This failure forces a silent parsing error for \texttt{y}, localizing the root cause to improper input handling rather than downstream logic. The robust version resolves this by utilizing \texttt{fgets} to capture full lines, initializing all variables, and explicitly verifying the parsed count.

\begin{lstlisting}[style=cstyle,caption={Delta Debugging exposes uninitialized input.},label={lst:example1}]
/* Fragile: only reads a single word from input */
char buf[16]; long long x, y;
scanf("%15s", buf);                // reads only up to first space
sscanf(buf, "%lld %lld", &x, &y);  // can set x, but y remains uninitialized

/* Robust: reads full line and validates number of inputs */
char buf[16] = ""; long long x = 0, y = 0;
if (fgets(buf, sizeof(buf), stdin) && 
    sscanf(buf, "%lld %lld", &x, &y) == 2) {
    /* ok: both x and y successfully parsed */}
\end{lstlisting}

Thus, delta debugging not only isolates faulty behaviors but also clarifies the nature of the failure, guiding LLMs toward structurally correct repairs. It complements test cases by converting them into minimal examples that sharpen fault localization and expose hidden assumptions in the code, revealing latent bugs that LLM-generated cases might miss.

\summary{\textbf{Summary 4:} Incorporating test cases into code repair improves performance, though the marginal benefit diminishes with more attempts, where excessive iterations can even degrade correctness. Integrating delta debugging successfully sustains this repair capability, especially for complex cases where test-feedback alone proves insufficient for LLMs.}

\subsection{RQ5: How readable is the code generated by \tool compared to rule‑based approaches?}

\begin{table}[ht]
\centering
\footnotesize
\setlength{\tabcolsep}{5pt}
\caption{Comparison of Code Readability in \tool Using Different LLMs and a Rule-Based Approach.}
\label{tab:readability}
\begin{tabular}{llllllll}
\toprule
 Dataset& & Ground-Truth & Gemma-3 & GPT-4o-mini & Llama-3.3 & Qwen-2.5-coder & GnuCOBOL\\
\midrule
\multirow{2}{*}{Subset} & Human & 4.35 & 3.78 & 4.05 & 3.85 & 3.83 & 1.34 \\
& LLM & 3.40 & 4.20 & 4.20 & 4.00 & 4.30 & 1.70 \\
\midrule
Entire dataset & LLM & 3.60 & 3.17 & 3.58 & 3.69 & 3.59 & 1.47 \\

\bottomrule
\end{tabular}
\end{table}

To evaluate readability, we conduct both human and automated experiments. For the human evaluation, we select 10 test cases where the code translated by all LLMs achieves a passing rate exceeding 97\%. Twelve computer science students (2 undergraduates, 5 Master's, 5 PhDs) rate the readability of the translated and ground-truth C programs on a 5-point Likert scale (1: extremely poor, 5: excellent), following established guidelines~\cite{BuseW10, PosnettHD11, abs-2407-03790}. For the automated evaluation, we employ Grok Fast 1 with the same prompt guidelines to assess code readability. The complete evaluation guidelines are available in our replication package's appendix.

Table~\ref{tab:readability} summarizes the results of both automated and human evaluations. \revise{In the human evaluation, the ground-truth code (original human-written code provided in the CodeNet dataset) achieves the highest readability score, closely followed by \tool's translations.} Conversely, the rule-based approach scores significantly lower than both. For the automated LLM evaluation, despite minor scoring inconsistencies with human judgment (e.g., Qwen-2.5-coder receiving the highest score), the overarching trend remains identical. Across the entire dataset, the LLM evaluation confirms that human-written and \tool-translated code exhibit highly comparable readability, both substantially outperforming the rule-based baseline. These findings demonstrate that while rule-based translations suffer from poor readability, \tool produces highly human-readable code, confirming its effectiveness.

\begin{figure}[ht]
\centering
\includegraphics[width=\linewidth]{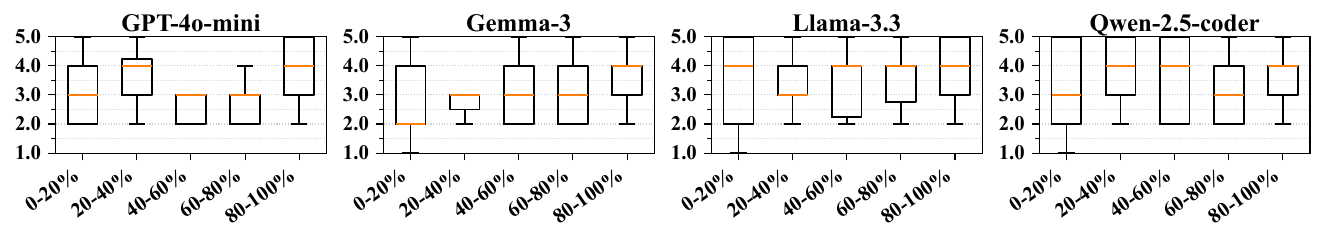}
\caption{Correlation between translated code correctness and readability (The X-axis denotes the pass rate of golden test suites, while the Y-axis indicates readability scores).}
\label{fig:readability_correct}
\end{figure}

To explore the relationship between readability and functional correctness, we perform a stratified analysis of readability scores across 20-percentage-point intervals of test case passing rates. Intuitively, a positive correlation might be expected, where higher functional correctness corresponds to superior code readability. However, as illustrated in Figure~\ref{fig:readability_correct}, readability scores exhibit no apparent trend across varying levels of correctness. These results indicate that the readability of LLM-translated code is largely independent of its functional correctness.

\summary{\textbf{Summary 5:} \tool translates COBOL into C with readability close to human-written code and significantly superior to rule-based approaches. Moreover, our evaluation demonstrates that the readability of LLM-translated code is largely independent of its functional correctness.}

\section{Discussion}

\revise{In this section, we analyze the specific COBOL code structures—including fixed-width record layout, slicing, loop bounds, and output reconstruction—that may lead to translation failures.}

\subsection{Failed translation case of COBOL code with fixed-width record layout}

\revise{Listing~\ref{cob_1} shows the COBOL code with a fixed-width record layout, and Listing~\ref{c_1} presents the corresponding translated C code. In this example, the COBOL code defines grouped input records with \code{FILLER} fields that encode fixed separator positions within the record layout. The \code{ACCEPT INP1} statement reads data directly into this grouped structure. However, LLMs employ a simple char array of length three and read input via whitespace-separated scanning, which also differs from COBOL’s positional layout.}

\begin{lstlisting}[style=cobolstyle, caption={COBOL code with fixed-width record layout}, label={cob_1}]
01 INP1.
    03 N PIC 9.
    03 FILLER PIC X.
    03 M PIC 9.
...
ACCEPT INP1.          
PERFORM VARYING I FROM 1 BY 1 UNTIL I > M
    ACCEPT INP2       
END-PERFORM.
\end{lstlisting}

\begin{lstlisting}[style=cstyle, caption={Translated C code with fixed-width record layout}, label={c_1}]
int main() {
    char inp1[3];    
    ...
    scanf("%1d %1d", &n, &m); 

    for (i = 1; i <= m; i++) {
        scanf("%1d %1d", &s, &c);
        ...
    }
}
\end{lstlisting}

\subsection{Failed translation case of COBOL code with slicing, loop bounds, and output reconstruction}

\revise{Listing~\ref{cob_3} showcases several COBOL-specific semantic conventions, whose incorrect C translations are shown in Listing~\ref{c_3}.  First, the substring expression \code{S(ini:len2)} relies on 1-based indexing, but is translated as \code{S + ini} instead of the required \code{S + (ini - 1)}. Second, the loop \code{PERFORM VARYING i FROM l BY -1 UNTIL i = 1} is incorrectly mapped to \code{for (i = l; i >= 1; i--)}. Because COBOL assumes \code{TEST BEFORE} semantics, the condition must be checked before execution, meaning the iteration for $i = 1$ should not run.  Third, the translation collapses COBOL's complex field-level rendering and input/output handling (\code{ACCEPT}, \code{UNSTRING}, \code{STORED-CHAR-LENGTH}) into ordinary C string operations (\code{fgets}, \code{strlen}, \code{sscanf}). Specifically, the original logic splits a field via \code{UNSTRING} ... \code{DELIMITED BY ALL SPACE} for output reconstruction, whereas the generated C code improperly formats and prints a null-terminated string directly.}

\begin{lstlisting}[style=cobolstyle, caption={COBOL code with slicing, loop bounds, and output reconstruction}, label={cob_3}]
COMPUTE len = FUNCTION STORED-CHAR-LENGTH(S).
...
PERFORM VARYING i FROM l BY -1 UNTIL i = 1
    ...
    IF S(1:len2) NOT = S(ini:len2) THEN ... END-IF
END-PERFORM.

MOVE i TO ZS.
PERFORM UNANS.
DISPLAY ans(1:FUNCTION STORED-CHAR-LENGTH(ans)). 

UNANS SECTION.
    UNSTRING ZS DELIMITED BY ALL SPACE INTO DUMMY ans END-UNSTRING.
\end{lstlisting}

\begin{lstlisting}[style=cstyle, caption={Translated C code with slicing, loop bounds, and output reconstruction}, label={c_3}]

int main() {
    ...
    for (i = l; i >= 1; i--) { 
        ...
        if (strncmp(S, S + ini, len2) != 0) { 
            flg = 0;
        }
        ... 
    }

    snprintf(ZS, sizeof(ZS), "%d", i);

    sscanf(ZS, "%s", ans); 
    printf("%s\n", ans);
    return 0;
}
\end{lstlisting}
\section{Related Work}
\label{sec:relatedwork}

\subsection{Automated Code Translation}
Research on code migration has progressed from manual rewriting and rule-based transcompilers to modern machine learning. Early methods leveraged foreign function interfaces like SWIG~\cite{beazley1996swig} or source-to-source transpilers like Emscripten~\cite{zakai2011emscripten}. Subsequent approaches applied statistical machine translation~\cite{karaivanov2014pbstatmt}, repository mining~\cite{9402018}, and unsupervised Transformer models like TransCoder~\cite{roziere2020unsupervised}. Recently, Pan et~al.~\cite{Pan_2024} introduced iterative repair via compiler and test feedback, which UniTrans~\cite{unitrans} generalized across languages using LLM-generated tests. For COBOL translation, Gandhi et~al.~\cite{gandhi2024translation} proposed a three-phase refinement strategy. Other advancements explore reinforcement learning~\cite{sakharova2025integrating, jana2023cotran}, repository-level scaling~\cite{ibrahimzada2025alphatrans}, agentic workflows~\cite{yuan2024transagent, luo2025unlocking}, and semantic reasoning~\cite{neurips_semcoder}. Unlike these code-level methods, alternative domain-level approaches generate functional descriptions~\cite{leveragingllm} or leverage intermediate representations~\cite{cobolformalmethods, umlocltrans} to guide the translation, whereas \tool operates directly at the code level.

\subsection{Using LLMs for Legacy Code}
Legacy programming languages introduce unique challenges due to archaic syntax and hidden semantics. Diggs et~al.~\cite{diggs2025leveraging} used LLMs to generate comments for MUMPS and ALC, while Ranasinghe et~al.~\cite{ranasinghe2025llm} and Chen et~al.~\cite{chen2024fortran2cpp} achieved promising results in FORTRAN-to-C++ translation via fine-tuning. To manage complexity, Luo et~al.~\cite{luo2025unlocking} and Macedo et~al.~\cite{macedo2024intertrans} proposed intermediate representation pipelines, whereas Lei et~al.~\cite{lei2025enhancing} used agentic methods for documentation. Additional studies focus on domain-specific rewrites with functional equivalence proofs~\cite{neurips_verifiedtranspllm} and architectural modernization~\cite{vikramusingai} to ensure correctness at system scale. Such migration necessitates automated testing; search-based software testing (SBST) tools like Pynguin~\cite{lukasczyk2022pynguin} implement MOSA~\cite{mosa} and DynaMOSA~\cite{dynamosa} strategies. Hybrid approaches combine SBST with LLMs, including CodaMOSA~\cite{lemieux2023codamosa} and TELPA~\cite{yang2024enhancing} for LLM-guided evolutionary testing, MuTAP for mutation testing~\cite{dakhel2024effective}, and CoverUp for iterative unit test generation~\cite{altmayer2025coverup}. Furthermore, symbolic execution tools like KLEE~\cite{cadar2008klee} and SymCC~\cite{poeplau2020symbolic} systematically explore execution paths to uncover corner cases~\cite{bailey2025symbolic}, \revise{with platforms like UTBot easing industrial application~\cite{utbotpresentation}, extending to more languages via fuzzing~\cite{utbotpython}, or mitigating path explosion via machine learning~\cite{learch}}. Iterative LLM-based repair pipelines have also proven effective in resolving recurring error patterns~\cite{fan2023automated, Pan_2024}.

Building on these approaches, we combine symbolic execution with LLM-generated tests within a repair loop, apply delta debugging to minimize failing cases into counterexamples, and embed COBOL semantics in prompts for both functional correctness and readability. While readability is challenging to measure automatically due to metric limitations~\cite{readabilitymetricmismatch}, empirical guidelines like reduced nesting~\cite{lessnestinghelps} correlate with human perception and guide our evaluation. Additionally, \tool's outputs are evaluated using LLMs for broader quality assessment, aligning with existing COBOL translation workflows~\cite{qualityevalcobolllm} and general code evaluation practices~\cite{zhuo-2024-ice,tong2024codejudge}.

\section{Threats to Validity}
Although our empirical study demonstrates substantial improvements over state-of-the-art baselines, several threats may limit the generalizability of our results.

\paragraph{Accuracy gap with rule-based tools.} \revise{While \tool outperforms pure LLM baselines, its accuracy still lags behind mature rule-based COBOL translation tools that utilize engineered grammatical constraints. Consequently, our LLM-centric framework retains inherent limitations in fully matching the precision of industrial-grade solutions.}

\paragraph{Evaluation data and scale.} We evaluate \tool on 319 function-level COBOL programs from IBM’s CodeNet, which cover only a small fraction of real-world production code. Furthermore, our function-level approach does not handle complex repository-level file interactions, meaning performance on larger industrial scales remains to be verified.

\paragraph{Baselines and configurations.} For reproducibility, we re-implemented UniTrans and HRJR using consistent prompts and deterministic decoding. However, alternative prompt templates, parameter settings (e.g., temperature sampling), or aggressive search strategies beyond our fixed repair iterations might yield different outcomes. \revise{Variations in LLM selection and deployment environments may also impact performance.}

\paragraph{Potential data leakage.} The evaluated COBOL programs might overlap with the proprietary training data of commercial LLMs. Due to the closed-source nature of these models and datasets, the exact extent of such performance inflation through memorization cannot be fully verified.

\paragraph{Target translation language selection.} We select C as the target language because mature open-source COBOL compilers predominantly target C, and COBOL-to-C migration aligns with critical industrial requirements (e.g., AWS's legacy modernization plans~\cite{aws_serverless_cobol}). Notably, while \tool's underlying transcompiler can be interchanged to target other languages, performance may vary.

\section{Conclusion}
We introduced \tool, a framework that translates COBOL into C by combining large language models (LLMs) with symbolic execution, automated test generation, and iterative repair. To handle COBOL’s unique syntax and semantics, \tool leverages delta debugging to simplify failing test cases, guiding the LLM toward accurate fixes. Our comprehensive evaluation demonstrates that \tool consistently outperforms state-of-the-art baselines by at least 12\%. These results showcase the effectiveness of integrating LLMs with symbolic and automated debugging techniques, providing a promising direction for legacy system modernization and future industrial applications.

\section{acknowledgment}
This paper was written as part of the Elite Graduate Program in Software Engineering offered by the University of Augsburg, Technical University of Munich and Ludwig-Maximilian-University of Munich.

\bibliographystyle{ACM-Reference-Format}
\bibliography{sample-base}


\end{document}